\documentclass[twocolumn,showpacs,preprintnumbers,prb]{revtex4}
\usepackage{graphicx}
\usepackage{dcolumn}
\usepackage{amssymb}
\usepackage{bm}

\begin{document}

\title{Interface and Phase Transition between Moore-Read and Halperin 331 Fractional Quantum Hall States: Realization of Chiral Majorana Fermion}
\author{Kun Yang}
\affiliation{National High Magnetic Field Laboratory and Department of Physics,
Florida State University, Tallahassee, Florida 32306, USA}
\pacs{73.43.Nq, 73.43.-f}

\begin{abstract}
We consider an interface separating the Moore-Read state and Halperin 331 state in a half filled Landau level, which can be realized in a double quantum well system with varying inter-well tunneling and/or interaction strength. We find in the presence of electron tunneling and strong Coulomb interaction across the interface, all charge modes localize and the only propagating mode left is a chiral Majorana fermion mode. Methods to probe this neutral mode are proposed. Quantum phase transition between the Moore-Read and Halperin 331 states is described by a network of such Majorana fermion modes. In addition to a direct transition, they may also be separated by a phase in which the Majorana fermions are delocalized, realizing an incompressible state which exhibits quantum Hall charge transport and bulk heat conduction.

\end{abstract}

\date{\today}

\maketitle

Fractional quantum Hall (FQH) states formed by electrons half-filling a Landau level (LL) have been of strong interest, due to the possibility that such states are non-Abelian. Prominent examples include the single-layer state at LL filling factor $\nu=5/2$,\cite{willett87} and states at $\nu=1/2$ formed in double\cite{eisenstein} or wide quantum well systems.\cite{suen} Most of the recent experimental efforts focused on the former, which (until very recently) was widely expected to be either the (non-Abelian) Moore-Read (MR, also referred to as Pfaffian),\cite{moore91} or its particle-hole conjugate, anti-Pfaffian state.\cite{levin07,lee07} Early interferometry experiments by Willett and co-workers\cite{Willett10} found interference patterns that are suggestive of braiding properties of non-Abelian quasiparticles, but this interpretation has been disputed.\cite{simon} Thermopower measurement,\cite{Chickering} which probes another aspect of non-Abelian quasiparticles, namely their quantum dimensions, turned out to be roughly consistent with theory,\cite{yanghalperin} but not yet reaching the accuracy that would allow for a definitive statement.
A review of experimental situation up to 2014 that includes experiments {\em not} directly probing the non-Abelian nature of the 5/2 state can be found in Ref. \onlinecite{lin}.
A very recent experiment\cite{Banerjee2} found a thermal Hall conductance on the 5/2 plateau that is consistent with neither the MR nor the anti-Pfaffian state.
Further complicating the situation is the fact that there exist many possible FQH states in a half-filled LL, some of which are actually Abelian. Energetically these states are often found to be competitive, based on numerical studies. This makes it difficult to pin-down the specific state that is realized in a given system. As a result at this point we do not have a system in which widely-accepted, {\em definitive} evidence for a non-Abelian state exists.

On the other hand the existence of a plural of competing FQH states in a half-filled LL offers the possibility of realizing multiple states in a single system separated by interfaces, as well as quantum phase transitions between such competing states. In the present paper we study a specific system in which such a possibility exists, namely in a double or wide quantum well system at half-filling,\cite{eisenstein,suen} the system can support either the (Abelian) Halperin 331 state,\cite{halperin} or the MR state. In our opinion these systems deserve considerably more experimental and theoretical attention than they have received. A recent numerical study\cite{zhu} (that builds on earlier work\cite{papic}) has shown convincingly that depending on system parameters, in particular strength of inter-well tunneling, both states can be realized. Due to the great control experimentalists have on such parameters, it is possible to create an interface between the 331 and MR states, if the numerical results turn out to be correct. We will show below that under fairly generic situations, the only gapless mode propagating along such an interface is a neutral chiral Majorana fermion mode. Detecting this mode experimentally will provide smoking-gun evidence that one of the two states involved is non-Abelian. We further consider the quantum phase transition between these two states in a disordered system, and show that it is described by a random network of Majorana wires.  We also note that such chiral Majorana fermion mode is of strong current interest.\cite{he}

\begin{figure}
\includegraphics[width=8cm]{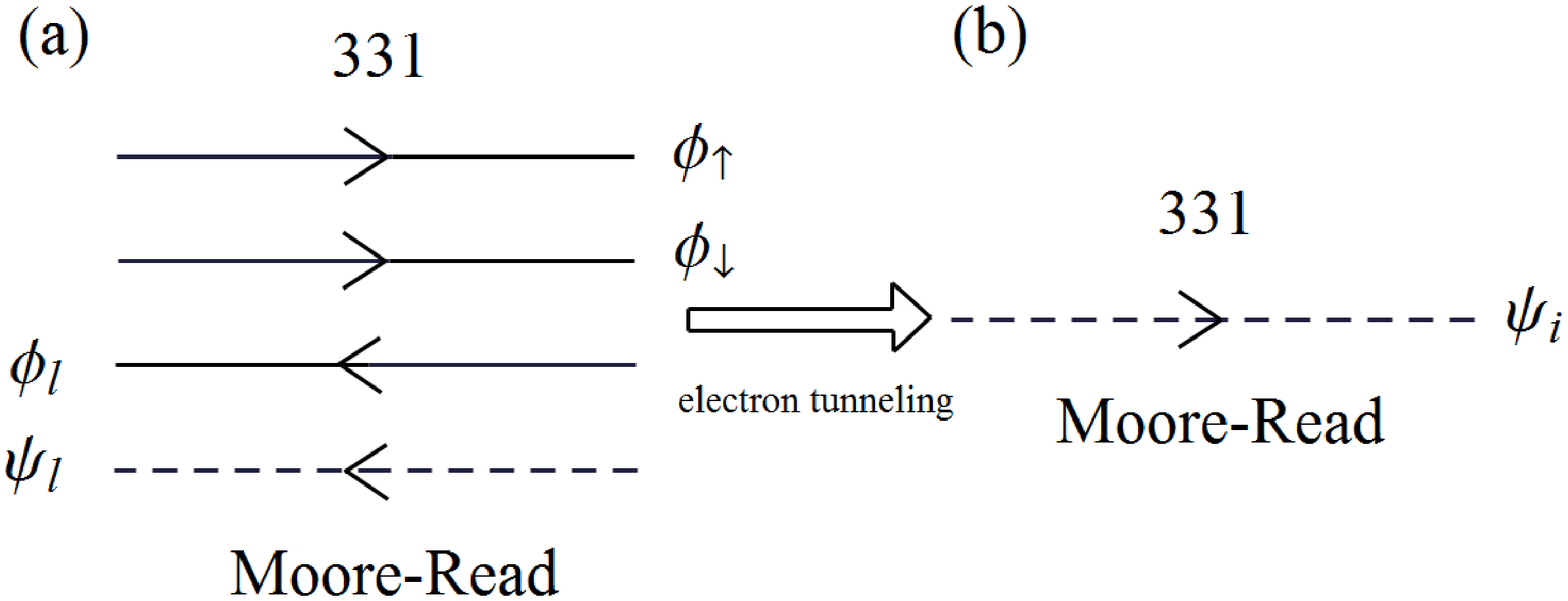}
\caption{Illustration of modes localized at the interface separating the 331 and Moore-Read states. (a): The two right-moving bosonic modes, $\phi_\uparrow$ and $\phi_\downarrow$, are the edge modes of the 331 state, corresponding to the upper and lower quantum well respectively. $\phi_l$ and $\psi_l$ are the left-moving bosonic and Majorana fermion edge modes of the MR state. Electron tunneling between the 331 and MR states mix these modes up and localize most of them, resulting in a single propagating neutral Majorana fermion mode (b). Majorana fermion modes are represented using dashed lines to highlight two differences with the bosonic modes (solid lines): (i) They are neutral while the bosonic modes are charged; and (ii) their central charge is 1/2, while the bosonic modes have central charge 1.}
\label{fig:domainwall}
\end{figure}

Topologically protected gapless edge modes of the 331 and MR states are well-understood.\cite{wen,milovanovic96} The degrees of freedom associated with an interface between them (see Fig. \ref{fig:domainwall}) is simply a combination of these modes aligned in opposite directions, described by the Lagrangian density
\begin{eqnarray}
L&=&\frac{1}{4\pi}\left(2\partial_t\phi_l \partial_x\phi_l -\sum_{ij} K_{ij}\partial_t\phi_i \partial_x\phi_j\right)  -i\psi_l\partial_t\psi_l \nonumber\\
 &-& H(\phi,\psi),
 \label{eq:L}
\end{eqnarray}
where $i,j=\uparrow,\downarrow$ are well indices, $\phi_{\uparrow,\downarrow}$ are the (right-moving) bosonic edge modes of the 331 state for upper and lower wells, and the $K$ matrix associated with the 331 state is
\begin{equation}
K = \left(\begin{array}{cc}
   3\; 1\\
   1\; 3\end{array}\right).
\end{equation}
$\phi_l$ is the left-moving bosonic mode of the MR state, and $\psi_l$ is its left-moving Majorana fermion mode, which is neutral. All three bosonic modes are charged, properly normalized such that the (one-dimensional) electron density along the interface is
\begin{eqnarray}
\rho(x)=\frac{1}{2\pi} \partial_x(\phi_l+\phi_\uparrow + \phi_\downarrow)=\frac{1}{2\pi} \partial_x\phi_c(x),
\end{eqnarray}
where
\begin{eqnarray}
\phi_c=\phi_l+\phi_\uparrow + \phi_\downarrow
\end{eqnarray}
is the total charge field. In anticipation to later usage, we introduce two other combinations of the bosonic fields:
\begin{eqnarray}
\phi_{r,n}=\phi_\uparrow \pm \phi_\downarrow,
\end{eqnarray}
where $\phi_{r}$ is the right-moving charge field associated with the 331 edge, and $\phi_{n}$ is a right-moving {\em neutral} field of the 331 edge. With these we also find \begin{eqnarray}
\phi_c=\phi_l+\phi_r,
\end{eqnarray}
namely the total charge field is the sum of the left- and right-moving charge fields.

In terms of these fields the dynamical terms (those involving time-derivatives) in Eq. (\ref{eq:L}) take the form
\begin{eqnarray}
\label{eq:dynamical terms}
L_0=\frac{1}{2\pi}\left(\partial_t\phi_l \partial_x\phi_l - \partial_t\phi_r \partial_x\phi_r \right) - \frac{1}{4\pi} \partial_t\phi_n \partial_x\phi_n   -i\psi_l\partial_t\psi_l.
\end{eqnarray}
A couple of comments are now in order. (i) The first term above corresponds to that of a {\em non-chiral} Luttinger liquid, with left- and right-mover described by $\phi_l$ and $\phi_r$ respectively. (ii) The second term above corresponds to that of a chiral fermion, which is the same as that of a right-moving $\nu=1$ quantum Hall edge (albeit the mode is neutral here). In particular, $\Psi_n(x)\sim e^{i\phi_n(x)}$ is a (neutral) fermion creation operator, or complex (Dirac) fermion field. Its real and imaginary parts, $\psi_r(x)\sim \cos[\phi_n(x)]$ and $\psi_i(x)\sim \sin[\phi_n(x)]$, are therefore real or Majorana fermion fields. We can thus now fermionize $\phi_n$ and rewrite $L_0$ as
\begin{eqnarray}
\label{eq:dynamical terms fermionic}
L_0&=&\frac{1}{2\pi}\left(\partial_t\phi_l \partial_x\phi_l - \partial_t\phi_r \partial_x\phi_r \right) -i(\psi_l\partial_t\psi_l - \psi_r\partial_t\psi_r) \nonumber\\ &+&i\psi_i\partial_t\psi_i,
\end{eqnarray}
where we have grouped the terms into a pair of boson fields with opposite chiralities, a pair of Majorana fermion fields with opposite chiralities, and a remaining right-moving chiral Majorana fermion field ($\psi_i$). In the above the subscript of $\psi_r$ can be understood either for real, or right-moving.

The Hamiltonian density $H(\phi,\psi)$ contains terms that correspond to all possible couplings among the modes of 331 and MR edges states, that are allowed by symmetry. We separate it into $H=H_0 + H'$, where $H_0$ is quadratic (and scale-invariant) in $\phi$ or $\psi$. The most general form of $H_0$ is
\begin{eqnarray}
H_0=\frac{1}{2\pi}\sum_{ab}v_{ab}(\partial_x\phi_a)(\partial_x\phi_b)   + \sum_{\alpha\beta}u_{\alpha\beta}\psi_\alpha\partial_x\psi_\beta,
\end{eqnarray}
where $a$ and $b$ take values $l$ or $r$, while $\alpha,\beta$ take values $i$, $l$ or $r$. $H_0$ describes a set of linearly dispersing modes that propagate along the interface. It is thus a fixed point Hamiltonian under renormalization group (RG), with dynamical exponent $z=1$.

Among terms in $H'$, the most relevant in the RG sense is random electron tunneling between 331 and MR edges, of the form
\begin{eqnarray}
H_T(x)&=& \xi(x) \psi_l(x) e^{i[2\phi_l(x)+3\phi_\uparrow(x)+\phi_\downarrow(x)]} + h.c.\nonumber\\
&+& \xi(x) \psi_l(x) e^{i[2\phi_l(x)+3\phi_\downarrow(x)+\phi_\uparrow(x)]} + h.c.\nonumber\\
&=& |\xi(x)| \psi_l(x) \cos[2\phi_c(x) + \varphi(x)] \cos[\phi_n(x)]\nonumber\\
&=& |\xi(x)| \psi_l(x)\psi_r(x) \cos[2\phi_c(x) + \varphi(x)],
\end{eqnarray}
where $\xi(x)$ is a random tunneling amplitude and $\varphi(x)$ is its phase.\cite{footnote} The first and second lines above correspond to tunneling between MR state and the upper and lower wells of the 331 state respectively. We assume short-range correlation in the tunneling amplitude:
\begin{eqnarray}
\overline{\xi^*(x) \xi(x')} = W\delta(x-x'),
\end{eqnarray}
where $W$ parameterizes the tunneling strength. $H_T$ describes combined backscattering between the two counter-propagating Majorana fermion modes, and the two counter-propagating bosonic charge modes. Such a term is allowed as we expect disorder to be inevitable at an interface, which is introduced by difference in sample properties on the two sides.

The nature of the propagating modes along the interface at large distance depends on the way $W$ scales under RG\cite{gs}:
\begin{eqnarray}
\frac{dW}{d\log s} = (3-2\Delta) W,
\end{eqnarray}
where $s$ is the length and time rescaling factor,
\begin{eqnarray}
\Delta =\Delta_{\psi_l} + \Delta_{\psi_r} + \Delta_c
\end{eqnarray}
is the scaling dimension of the operator $\psi_l(x)\psi_r(x) \cos[2\phi_c(x)]$, $\Delta_{\psi_l} = \Delta_{\psi_r} =1/2$ are the scaling dimensions of the free Majorana fields $\psi_l$ and $\psi_r$ respectively, and $\Delta_c$ is the scaling dimension of $\cos[2\phi_c(x)]$ which is non-universal and depends on $v_{ab}$.
If $\Delta > 3/2$, $W$ scales to zero at large distance and we expect all the modes described by $H_0$ remain propagating modes, with possibly renormalized velocities, at least for sufficiently small initial value of $W$. On the other hand if $\Delta  < 3/2$, $W$ increases indefinitely at large distance, and we expect $\psi_l, \psi_r, \phi_l$ and $\phi_r$ to be localized by strong backscattering; as a result the only propagating mode at large distance will be $\psi_i$, a right-moving Majorana fermion mode. Which one of these two situations gets realized depends crucially on the value of $\Delta_c$.

A straightforward calculation yields
\begin{eqnarray}
\Delta_c=4\sqrt{\frac{v_{ll}+v_{rr}-2v_{lr}}{v_{ll}+v_{rr}+2v_{lr}}}.
\end{eqnarray}
The dominant contribution to $v_{ab}$ comes from the Coulomb interaction at the interface of the form $v_c(\partial_x\phi_c)^2$, which by itself yields $v_{ll}+v_{rr}-2v_{lr}=0$. In the absence of screening $v_c$ diverges logarithmically in the large distance limit. In reality due to metallic gates nearby, screening is present, but only at distance larger than the spacing between the 2D electron gas and these gates. We thus expect $v_c \gg v'$, where $v'$ represents other contributions to $v_{ab}$ whose form depends on details of the interface. As a result we expect $\Delta_c \sim O(\sqrt{v'/v_c}) \ll 1$, rendering $\Delta  < 3/2$. Physically this simply reflects the fact that strong Coulomb interaction suppresses the fluctuation of total charge field $\phi_c$, leading to low scaling dimensions of vertex operators of $\phi_c$. As a result backscattering is {\em relevant}, and all charge modes are localized.

There are some similarities and differences between what happens here and at the interface between a MR state and its particle-hole conjugate, anti-Pfaffian state.\cite{wy} There the charged modes are gapped out at the interface, due to correlated tunneling of a pair of electrons, leaving only propagating neutral modes at long-distance/low-energy.
Disorder, however, was not needed there as the electron pair tunneling conserves momentum, due to particle-hole symmetry at half-filling, and the interface may form {\em spontaneously} in an otherwise homogeneous sample.
The single electron tunneling that drives the localization here, on the other hand, does {\em not} conserve momentum, and is allowed only in the presence of disorder.

Chiral Majorana fermions are of strong current interest. They can live at the edge of a $p_x + i p_y$ topological superconductor, and are argued to be responsible for the recently observed half-integer quantization of two-terminal conductance in a quantum anomalous Hall/$s$-wave superconductor hybrid system.\cite{he} However alternative interpretations of the experiment were offered,\cite{ji,huang} and the topic remains controversial.\cite{lian} Our result indicates the 331/MR interface is another system that supports a stand-alone chiral Majorana fermion mode. In the following we discuss possible experimental probes of its existence.

\begin{figure}
\includegraphics[width=8cm]{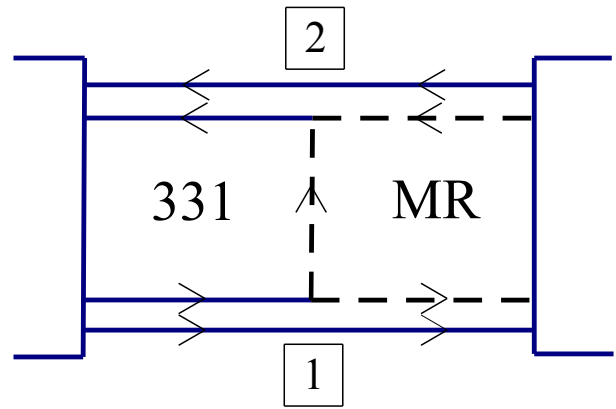}
\caption{Illustration of two-terminal charge and heat transport across or along an MR/331 interface. Solid lines represent charged bosonic modes, and dashed lines represent neutral Majorana fermion modes, with arrows indicating their propagation direction. The boxes labelled 1 and 2 are heat sources/sensors, that can be used to inject/detect heat into the edge or interface modes. Here it is assumed that the contacts are perfectly lined up with the interface, which requires the precise control that might be overly optimistic for the experimentalists. In reality it is likely that tweaks of the proposed setup are necessary, including having more thermal contacts than simply the two illustrated here.}
\label{fig:two-terminal}
\end{figure}

Consider a simple configuration illustrated in Fig. \ref{fig:two-terminal}. The two-terminal conductance of the MR and 331 states are both $G=\frac{e^2}{2h}$. By controlling, {\em e.g.}, the potential barrier height separating the two quantum wells, one can prepare a sample that favors 331 state on the left and MR state on the right, with an interface in the middle. Based on the theory above, there is a neutral Majorana fermion mode connecting the two (current carrying) edges. Due to the neutrality of this mode, it does {\em not} induce back scattering of electric current; as a result we still have $G=\frac{e^2}{2h}$, in the zero temperature limit. On the other hand if there were charged modes propagating along the interface, we expect back scattering and thus $G<\frac{e^2}{2h}$.

Finding $G=\frac{e^2}{2h}$, is, of course, a null result that could simply be due to the absence of the interface. To study the neutral interface mode we need to rely on thermal transport, as the only conserved quantity it can carry is energy. To probe the chirality of the mode we can send in a heat pulse through thermal contact 1, which will propagate along the interface, resulting in a thermal signal in contact 2. On the other hand a heat pulse in contact 2 does {\em not} generate a response in contact 1. This difference unambiguously determines the chirality of the interface mode. More quantitatively, the heat carrying ability of the interface mode is characterized by its thermal conductance
\begin{eqnarray}
G_Q = c\kappa_0 T,
\end{eqnarray}
where $T$ is temperature, $\kappa_0=\pi^2 k_B^2/(3h)$ is a fundamental constant that sets the upper bound of the heat carrying ability of a ballistic 1D channel,\cite{pendry} and $c$ is the central charge of the interface mode which is $1/2$ here. By measuring $G_Q$ using the configuration of Fig. \ref{fig:two-terminal} or its generalizations one can directly determine $c$, which allows for an unambiguous identification of the chiral Majorana fermion mode. We note (fractionally) quantized heat transport has been observed recently in the FQH regime,\cite{Banerjee1,Banerjee2} which justifies optimism of using this method to pin down a {\em stand-alone} chiral Majorana fermion mode we find here. The existence of such a mode in turn guarantees that at least one of the FQH phases is non-Abelian, as Abelian quantum Hall edge states give rise to {\em integer} quantized thermal transport.\cite{Banerjee1,KaneFisher} We also note that there is significant advantage of using such interface to probe bulk topological order as compared to the more widely used edge states, because the electrostatic environment at the interface is essentially identical to that of the bulk for the 2D electron gas. The situation is very different at the edge; there due to the termination of compensating background charge, edge reconstruction occurs rather generically which complicates the physics in a way that may mask the universal topological physics.\cite{edgeReconstructionPapers}

\begin{figure}
\includegraphics[width=8cm]{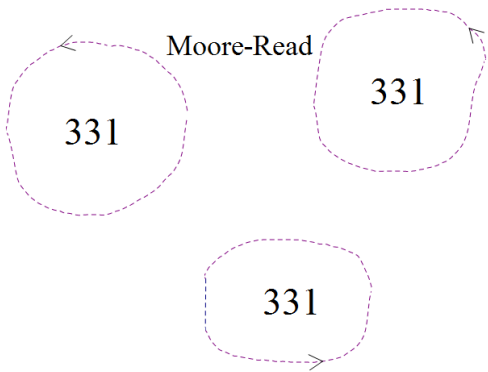}
\caption{Domains of 331 state in the Moore-Read back ground. Bulk low-energy degrees of freedom that drive the transition between them are the chiral Majorana fermion modes propagating along the domain walls.}
\label{fig:domains}
\end{figure}

We now consider the quantum phase transition between the MR and 331 states in a disordered system, driven by, {\em e.g.}, separation between the two quantum wells or tunneling strength between them. In a disordered system with inhomogeneity, we expect domains of both phases to coexist near the transition, and the transition to be that of the (quantum) percolation-type. The degrees of freedom that dominate low-energy physics and ultimately drive the transition are precisely the chiral Majorana fermion modes propagating along the domain walls (see Fig. \ref{fig:domains}), and the transition is nothing but the localization-delocalization transition of these Majorana fermions. The physics can be captured using a network model, with Majorana fermions propagating along its links. Such a network model was studied in Ref. \onlinecite{ChalkerRead}, where it was found that it belongs to class D in the symmetry classification of Altland and Zirbauer.\cite{AltlandZirbauer} As illustrated in Fig. 1 of Ref. \onlinecite{ChalkerRead}, the system supports three phases under the most generic situations; the MR and 331 states correspond to the two localized phases, with trivial and non-trivial topological quantum numbers. Interestingly, in addition to the possibility of a direct 2nd order transition between them, there can also be an intermediate ``metallic" phase separating them, in which the Majorana fermions are delocalized. Since these modes are neutral, in fact our system is incompressible and thus in the FQH phase as long as charge transport is concerned; the ``metallic" phase is actually a thermal metal in which the bulk is a heat or energy (but not charge) conductor. This is similar to the striped quantum Hall state discussed in Ref. \onlinecite{wy}, and such bulk heat conduction has been observed in a number of FQH states.\cite{inoue14}
This ``metallic" phase is quite unusual, because the Majorana fermion density of states is expected diverge logarithmically as one approaches zero energy.\cite{SenthilFisher} This results in logarithmical divergence in the linear $T$ coefficients of both thermal conductivity and specific heat at low temperature.\cite{SenthilFisher} Observation of such behavior will reveal the presence of such a ``metallic" state. We note there has been a recent report of specific heat measurement of 2D electron gas in the FQH regime.\cite{Gervais}

We now compare out results with an existing experiment.\cite{suen94} In this work the charge gap was measured as a function of increasing tunneling strength, and a sharp upward cusp (or maximum) was found. Such increasing tunneling is expected to drive the system from double-quantum well to single-quantum well regimes, and the cusp was recently interpreted\cite{zhu} as transition from the (double-quantum well) 331 to (single-quantum well) MR states. Indeed similar transitions were found in the original experiment\cite{suen94} at other filling factors like 2/3. Different from the 1/2 state however, other transitions manifest themselves as a {\em downward} cusp (or minimum) in charge gap, which is easy to understand as (low-energy) interface modes that proliferate near the transition can carry charge current. The opposite behavior at 1/2, on the other hand, indicate such interface modes or low-energy degrees of freedom near the transition (or in the intermediate phase if there is one) must be charge-neutral, consistent with our theory.

We note that in addition to double or wide quantum well systems, our results may also apply to quantum Hall states in single or multi-layered graphene,\cite{barlasreview} where fractional quantum Hall states at half-filling and quantum phase transitions involving them have been observed recently.\cite{young} Also of interest are interfaces between different non-Abelian quantum Hall states, like that studied in Ref. \onlinecite{grosfeld}; in particular, a network built on the Pfaffian/anti-Pfaffian interface studied in Refs. \onlinecite{wy} and \onlinecite{Barkeshli} have been shown very recently\cite{mross,wang} to support a plural of novel phases, including a thermal metal phase discussed here and in Ref. \onlinecite{wy}.

The author thanks Loyd Engel for helpful discussions, and Li Chen for assistance.
This work was supported by the National Science Foundation (Grants No. DMR-1157490 and No. DMR-1442366).


\begin{references}
\bibitem{footnote} Microscopically the tunneling amplitude $\xi(x)$ can be different for the upper and lower layers, as disorder at the interface {\em locally} breaks layer symmetry (that is present in a clean system). However the bosonized description we use here is an effective field theory that is appropriate at longer length scales, in which the microscopic degrees of freedom have been coarse-grained, as a result of which $\xi (x)$ has been averaged within this scale. Such averaging restores the layer symmetry, resulting in the same $\xi (x)$ for the two layers.
\end{references}
\end{document}